\documentclass[12pt,preprint]{aastex}

\shorttitle{The 3.5-h period in V382 Vel and the Long Term Light Curve}
\shortauthors{Balman \c{S}. et al.}                                                

\begin{document}
\title{The Detection of a 3.5-h Period in the Classical Nova Velorum 1999
(V382 Vel) and the Long Term Behavior of the Nova Light Curve 
}
 \author{ \c{S}\"olen Balman\altaffilmark{1,2}, Alon Retter\altaffilmark{3}, 
Marc Bos\altaffilmark{4}
\altaffiltext{1} {Middle East Technical University, In\"on\"u Bulvar{\i}, Ankara, Turkey, 06531; solen@astroa.physics.metu.edu.tr}
\altaffiltext{2} {Astrophysics Missions Division, Research and Scientific Support Department of
ESA, ESTEC, Postbus 299 SCI-SA, Keplerlaan 1 2201 AZ Noorwijk ZH, The Netherlands}
\altaffiltext{3} {Department of Astronomy and Astrophysics, Pennsylvania State 
University, 525 Davey Lab., University Park, PA 16802, USA}
\altaffiltext{4} {Mount Molehill Observatory, Auckland, NZ}}
\begin{abstract}

We present CCD photometry, light curve and time series analysis
of the classical nova V382 Vel (N Vel 1999). The source was observed
for 2 nights in 2000, 21 nights in 2001 and 7 nights in 2002 using clear filters.
We report the detection of a distinct period in the
light curve of the nova P=0.146126(18) d (3.5 h). 
The period is evident in all data sets,
and we interpret it as the binary period of the system.
We also measured an increase in the amplitude modulation of the optical light (in magnitude) by more than
55$\%$ from 2000 to 2001 and about 64$\%$ from 2001 to 2002.  
The pulse profiles in 2001 show deviations from a pure sinusoidal shape which progressively
become more  sinusoidal by 2002. The main cause of the variations in 2001 and 2002 can be
explained with the occultation of the accretion disk by the secondary star. 
We interpret the observed deviations from a pure
sinusoidal shape as additional flux resulting from the aspect variations of the irradiated 
face of the secondary star.
  
\end{abstract} 


\section{INTRODUCTION}

A classical nova outburst arises from the explosive ignition of
accreted matter
(i.e., thermo-nuclear runaway, TNR) in a cataclysmic binary system
where a Roche-lobe filling secondary transfers hydrogen rich
material typically via an
accretion disk onto the white dwarf (WD) primary.
During the outburst, the envelope of the WD expands to $\sim$ 100
R$_{\odot}$ and 10$^{-7}$ to 10$^{-3}$\ M$_{\odot}$ of material
that exceeds the escape velocity is expelled
from the system (Shara 1989; Warner 1995; Starrfield 2002).
The outburst stage usually lasts from a few
months to several years and finally, the system returns to a quiescent
state after the outburst.

Nova Vel 1999 (V382 Vel) was discovered on 1999, May 20.6 UT
(Lee at al. 1999). It is believed to be an
outburst on an ONeMg white dwarf because of the strong Ne line emission
detected in the optical wavelengths (Woodward et al. 1999).
Spectroscopic observations from 5 to 498 days (after the eruption) indicate that
the nova belongs to a broad Fe II spectroscopic class with an absolute magnitude
at the optical maximum M$_v$$\sim$-8.9, t$_2$=6 d and t$_3$=10 d
(Della Valle et al. 2002). The early evolution of the nova shows 
an iron curtain phase and P-Cygni profiles on all the important resonance lines
with expansion velocities between 2000-4000 km s$^{-1}$ (see Shore et al. 2003 and 
Della Valle et al. 2002 for the optical and Burwitz et al. 2002 
for the X-ray wavelengths).
Fragmentation in the ejecta is apparent and the ejected mass is calculated to be
4-5$\times$10$^{-4}$ M$_{\odot}$. The nova is
found to be  enhanced in Ne/Ne$_{\odot}$=17($\pm$3), N/N$_{\odot}$=17($\pm$4), 
C/C$_{\odot}$=0.6($\pm$0.3), Al/Al$_{\odot}$=21($\pm$2), and Mg/Mg$_{\odot}$=2.6($\pm$0.1) 
(Shore et al. 2003). The distance to the nova is estimated as 1.7-2.5 kpc (Della Valle et al. 2002; 
Shore et al. 2003).
In addition, Nova Vel 1999 appeared as a Super Soft X-ray
Source (Orio et al. 2002; Ness et al. 2005) and the hard X-ray emission 
was detected originating
from the shock heated material in the shell (Orio et al. 2001; Mukai $\&$ Ishida 2001; 
Burwitz et al. 2002; Ness et al. 2005)

Bos et al. (2002) discovered  a 3.5-h periodicity;
0.14615(1) d together with its second harmonic in the optical light curve of the nova. 
In this paper, we will elaborate
that report, discuss possible physical mechanisms for the variation and
try to understand the characteristics of the long term
light curve and the detected period.

\section{OBSERVATIONS}

We observed V382 Vel during 2 nights in 2000,  21 nights in 2001 and
7 nights in 2002 through clear filters. Table 1 displays the summary
of the observation schedule. The observations in 2000 were obtained with 
a 0.25 m (Meade) telescope
at the Wharemaru Observatory, New Zealand with a ST6 CCD Camera.
The photometry in 2001 and 2002 was carried out using
the 0.3 m telescope at the Molehill Observatory with the ST6b CCD Camera, Auckland, New Zealand.
The typical exposure times were 20 s in 2000, and 45-60 s in 2001 and  2002. Also, three nights
in January 2001 were obtained at the 0.75 m telescope of the Sutherland, 
South African Observatory using a UCT CCD photometer. 

Before carrying out the photometric measurements, standard noise reduction
was applied to the images and the bias and flat corrections were made.
Aperture photometry was performed on the corrected and normalized images. 
A reference group of  three comparison stars, close to the nova, in the same field
were used in order to reduce the scintillation effects and  derive the relative magnitudes.
In Figure 1, we indicate the times of our obseravtions with arrows in a plot of the light curve
obtained by the AAVSO covering data from the initial outburst to 2003.

\section{DATA ANALYSIS AND RESULTS}

Using our reduced and calibrated data,
we constructed light curves for the given nights in Table 1.
A collection of normalized light curves obtained from the
longer runs are  displayed on Figure 2 for 2000, 2001, and
2002. The observational dates are indicated on each panel of the figure.
A modulation of the light curve can  be seen in all nights. 
The typical errors of the data points (in magnitudes) are 0$^m$.03
for 2000, 0$^m$.014 for 2001 and 0$^m$.013 for 2002 as calculated from the
standard deviation of the magnitudes of the stars chosen as reference to 
calculate the differential magnitudes. 

We performed Fourier analysis on the  
time series obtained from the data in order to derive the period of the
modulations using the ESO-MIDAS (European Southern Observatory Munich Image Data Analysis System) 
time series analysis package. 
Several standard programs were used
like the Scargle Algorithim (Scargle 1982) and Discrete Fourier
Analysis using Leahy normalization (Leahy et al. 1983).
Figure 3, top panel, shows the power spectrum of the data
for the year 2000 where the Scargle algorithm was used for the analysis.
The middle and bottom panels in Figure 3 show the power spectrum for the years 2001 and 2002, 
respectively, derived using similar analysis techniques. 
The detection limit of a period at the 3$\sigma$ confidence level (99$\%$) is
a power of 18.2 in the middle panel (2001 data), 
a power of 13.8 in the top panel (2000 data) and a power of 15.8 in the bottom panel (2002 data)
(see also Scargle 1982).
Before calculating the power spectra, the individual or consecutive 
nights were normalized by subtraction of the mean magnitude. 
In order to correct for the effects of time windows and sampling on power spectra,
synthetic constant light curves were created and a  few very prominent frequency peaks that appear
in these light curves were prewhitened from the data in the analysis.
When necessary, the red noise in the lower frequencies
was removed by detrending the data using linear or quadratic fits.
There was considerable red noise in the power spectra especially for the years  2001 and 2002.
The red noise level at the low frequencies below 10 d$^{-1}$ (2$\times 10^{-4}$ Hz)
increased by a factor of three from 2000 to 2001 and 2002.

We found a prominent period at P=0.146126(18) d using the whole data set. 
The power spectra on Figure 3
show the highest peak at this period
and the group of peaks around
it are some of the $\pm$1/3, $\pm$1/2, $\pm$1, and $\pm$2 day 
aliases of the detected period.
In all the figures except Figure 3, top panel, the second harmonic of P is present.
We did not recover any other significant and/or persistent period in all the years (2000-2002).
The ephemeris for P determined by fitting a sine curve are :

\noindent
 T$_{0}$= HJD 2451966.820($\pm$0.001) + 0.146126($\pm$0.000018)E

The accumulated error on the period for the entire time span of our optical data is 
0.014 d (about 10$\%$ of one cycle).
Figure 4 displays the mean light curves folded on P. The top and bottom panels (among the three panels in the figure)
show the folded light curve of the 2001 and 2002 data sets, respectively. 
The short light curves obtained in 2000 have high statistical errors  
in comparison with the period modulation amplitude, which gives a   
large error on the detected period (see Figure 3, top panel). This, in return, yields a high ambiguity in the 
modulation amplitude and thus, we excluded the mean light curve from Figure 4.
The lines in the middle of the top and bottom panels on Figure 4 show the average variation in the differential
magnitude of the three reference stars used in the analysis folded on P 
(i.e., mean light curve of the reference stars). 
The number of bins in each panel (on Figure 4) are 
chosen in accordance with the accumulated error for each data set.  
The period P showed an amplitude variation of
$<$0$^{m}$.007 in 2000.
The amplitude of the variations were increased significantly
to 0$^{m}$.014$\pm$0$^{m}$.003 in 2001 and to 0$^{m}$.023$\pm$0$^{m}$.008 in 2002
where the shape was almost sinusoidal. 
The increase in modulation depth was more than  55$\%$ (in magnitude) from 2000 to 2001
and about 64$\%$ from 2001 to 2002 on the average. 

\section{DISCUSSION}

We presented 30 nights of data on V382 Vel obtained using 
clear filters in 2000, 2001 and 2002. We discovered modulations in the light curve
of the nova at the period P=0.146126(18) d with an amplitude of 
$\Delta m$$<$0.007 in 2000 and the amplitude
increased  to $\Delta m$= 0.014 in 2001 and to $\Delta m$=0.023 in 2002.
Since the periodicity is persistent and coherent, we propose that this is the 
binary period of the system. 

In general, eclipses or occultation effects are
detected from novae in outburst and in quiescence (Leibowitz et al. 1992; Kato et al. 2004;
Shafter et al. 1993, Woudt $\&$ Warner 2001, 2002, 2003). For example, 
V838 Her (N Her 1991) shows variable
eclipse depth that lasts from 2 to 3 hours, whose depth varies between 0.1 and 0.4 magnitues in about 4 month
(Leibowitz et al. 1992). The eclipse depth in V1494 Aql (N Aql No.2) is reported to change by 10 times
from 0.05 to 0.5 magnitudes in one year from July 2000 to July 2001 (Bos et al. 2001).

The increase in the modulation amplidutes detected throughout the 2000-2002 light curves indicates
an occultation in the binary system that is becoming more apparent with time.
This is also accompanied with the decreasing brightness/cooling of the WD by about 2 magnitudes
from 2000 to 2001 and by about 1.5 magnitudes after February 2001 until the end of January 2002
(see Figure 1).
We interpret the cause of the variations in 2001 and 2002 as the occultation of an accretion disk by the
secondary star.
Fading of the nova itself, is due to a decline in the optical radiation of an uneclipsed source.
Therefore, in time, the relative contribution of the occulted body to the total optical output of the 
system increases causing an increase in the observed amplitude of the variations.
Augusto $\&$ Diaz (2003)
found an increase in the blue continuum of the nova spectra by 565 days after the outburst  
and interpreted it as an indication of the
re-establishment of accretion and presence of an accretion disk in the system. This confirms our findings since our observations in 2001
start  about 645 days after the outburst.
The existence of the accretion disk (i.e., mass transfer via an accretion disk) is also supported by the 
increasing level of red noise seen in the power spectra in 2001 and 2002 as mentioned in sec.(3)
(see also van Der Klis 2000).
In addition, the expanding photosphere of the nova is already  below the
surface of the critical lobe of the WD by 2001 since the X-ray turn-off is between December 1999 and February 2000 (Ness et al. 2005).
Moreover, the nova/nova shell is already in the nebular stage by the beginning of our optical observations,
thus there can not have been any obscuration originating from the ejected material and/or circumbinary medium.      

After a nova explosion, the hot WDs may heat and irradiate their cooler companions
once they become Super Soft X-ray Sources (SSS) in the course of their evolution
(e.g, V1974 Cyg: DeYoung $\&$ Schmidt 1994; also, GQ Mus: Diaz et al. 1995; V1494 Aql:
Hachisu et al. 2004).
The orbital period of novae can be  detected as a result of
the aspect variations of the secondary due to heating from the hot WD
(Kovetz, Prialnik, $\&$ Shara 1988) and the asymmetry in the pulse profiles could be
produced once the shape of the secondary is of a tear drop model.
The irradiation effects in classical novae
can also be detected long after the outburst stage (e.g., V1500 Cyg: Sommers $\&$ Naylor
1999; DM Gem: Retter et al. 1999).
As mentioned before, V382 Vel was detected as an SSS
and found to turn off the H burning between December 1999 and February 2000 (about eight months
after the outburst).
Our observations start 
in 2000 May 30 and June 2 where the stellar remnant was still hot and the soft-flux 
was declining
in the X-ray wavelengths (Burwitz et al. 2002; Ness et al. 2005). 
As a result, we expect to see some irradiation
effects in this system.    

The orbital variations in V1974 Cyg were $\sim$0$^m$.1 in the $I$ band and
$<$0$^m$.05 in the $V$ band
about the time the H-burning turned-off in 1993 (the WD temperature was about
3-4$\times$ 10$^5$ K; Balman et al. 1998). The modulation depth in the I band decreased to about 0.025 magnitudes
in 1994 (one year later, Retter et al. 1997) 
where the remnant WD was no longer detectable in the X-ray wavelengths.  
Another classical nova detected as
an SSS showing irradiation effects, was GQ Mus (the WD temperature was about
5.1$\times$10$^5$ K in 1992;
Balman $\&$ Krautter 2001 and references therein).
The optical light curve showed modulation at the orbital period with an amplitude of 0.33 magnitudes in
1990 that decreased to 0.05 magnitudes in 1994 while the detected asymmetric profile of 1990 changed to
a symmetric flat-topped modulation in 1994 (Diaz $\&$ Steiner 1994; Diaz et al. 1995).
The optical pulse profile of GQ Mus in 1990 resembles the profile of V1974 Cyg in 1993 and
V2275 Cyg in 2003 (N Cyg 2001 No. 2) where the first two are novae with orbital periods below the period gap
and the latter a nova with an orbital period of about 7.6 hours (Balman et al. 2003; Balman et al. 2005).
The pulse shapes of the orbital modulation of V2275 Cyg changed significantly from 2002 to 2003 and 
the modulation amplitude decreased (at the orbital period) from 0.42 magnitudes in 2002 to 0.22 magnitudes in 2003, while
the irradiated face of the secondary cooled in time (Balman et al. 2005). This source was not observed
in the X-rays, thus an SSS phase is not confirmed.

A close inspection of the 2001 light curve of V382 Vel reveals that the modulation shape shows deviations from
a pure sinusoidal. Therefore, we constructed a synthetic sine wave using the orbital 
frequency and subtracted
this from the 2001 light curve which would be equivalent to the removal of the variations as a result of
the occultation of the disk. The middle panel of Figure 4 shows the residual emission
in the light curve of  V382 Vel  folded on the orbital period.
There is additional flux at phases about $\Phi$$\simeq$0.10-0.15 and 0.5-0.6. 
These phases correspond to the position of the secondary at about 90$^{\circ}$ to our line of sight. Thus,
these humps can possibly be explained 
by the emission from the irradiated side of the secondary (eg., at the vicinity of L1, the lagrangian point)  
falling into our line of sight.  The observed excess dip at about the orbital minimum 
(about $\Phi$$\simeq$0.8-0.9) is, then could be caused
by the self-occultation of the irradiated zone by the secondary star when the star is between the WD and 
the observer.
As time progresses, much less emission
from the irradiated secondary intervenes with the modulation depths (due to occultation of the accretion
disk) in 2002 resulting in a more sinusoidal pulse shape.
The scenario described here requires a relatively high inclination angle for the 
system. 
   
We did not recover any significant superhump period in the light curve of  V382 Vel given the time span of our 
observations that extends to about 985 days after outburst. Superhumps are QPOs caused by
tidal instabilities in accreting binary systems with M$_2$/M$_1$$<$0.35$\pm$0.02 (see Patterson et al. 
2005 for a general review; Nelemans 2005 for AM CVns). 
In classical novae superhumps can be used as probes to study how accretion is reestablished after the 
outburst. Several systems show such QPOs as early as two and a half months after the outburst
(V4633 Sgr, Lipkin et al. 2001) or 
even two years after the outburst like V1974 Cyg 
(Retter et al. 1997). 

We checked the possibility of any change in the detected binary period.
This could have revealed a system that was recoiling to its original size after the 
initial eruption with a slight expansion in the
binary separation. The entire light curve were separated into six consecutive observations and
the period was calculated for each set.                       
The data were consistent with a constant line and no period derivative.

\section{SUMMARY AND CONCLUSIONS}

We have detected  a consistent and coherent periodicity P=0.146126(18) in the optical light curve of the
classical nova V382 Vel using the data obtained in the years 2000 (two nights), 2001 (21 nights) and 2002 
(seven nights).
The observations were conducted at the Wharemaru Observatory (2000; 0.25m telescope) and Molehill Observatory
(2000-2001; 0.3m telescope) in New Zealand. We interpret this variation as the binary period of the underlying system.
We conclude that the cause of variations is the occultation of an accretion disk by the secondary star.
We detect increasing modulation amplitudes as the nova itself cools off in time  in the 
optical wavelengths, and the contribution of light from the occulted disk into the total
light curve increases enhancing the amount of variation in the light curve.
The increase in the variation amplitude from 2000 ($\Delta$m$<$0.007) to 2001 
($\Delta$m=.014$\pm$0.003) is larger than 55$\%$ and from 2001 to 2002 ($\Delta$m=0.023$\pm$0.008), it is 64$\%$.
We also observe variation of the shape and depth of the mean light curves particularly in 2001.
We favor a scenario where variations due to the aspect changes of the irradiated secondary star 
intervenes with the optical light curve at phases 
0.10-0.15, 0.5-0.6 and 0.8-0.9. 

The soft X-ray radiation from the WDs in the outburst stage of classical novae
could trigger irradiation induced high mass transfer resulting 
in re-establishment of the accretion disk. Irradiation induced 
mass transfer cycles
occur in compact binaries if the donor star has a shallow convective envelope (fast thermal time scale),
or the system has mass transfer driven on a long dynamical timescale, or the 
photosphere scale height is small (Buning $\&$ Ritter 2004; Ritter, Zhang, $\&$ Kolb 2000). 
Once the nova is a soft X-ray source it has the potential to 
irradiate its companion initiating mass transfer (stable/limit cycle)
if the basic conditions described above are also met depending on other important parameters
like the masses of the two stars, the binary separation, etc.
The systems which have orbital periods above 3 hrs
tend to form mass
transfer limit cycles whereas the systems with periods 
below 2.5 hrs show stable
mass transfer unless the deriving rate is no larger than a few times
the gravitational braking rate (Buning $\&$ Ritter 2004).
The H burning phase of classical novae outbursts will be interesting laboratories to
study irradiation induced mass transfer phenomenon for CVs. In addition, the hibernation scenario for novae
suggests that classical novae remain bright for a few centuries after the eruption
because of irradiation induced mass transfer (Shara et al. 1986; Prialnik
$\&$ Shara 1986). Nova systems are known to show 
irradiation effects and/or superhumps in their light curves after the optical decline   
(see Retter $\&$ Naylor 2000). 

The three years of monitoring observations  of the classical nova V382 Vel 
in the optical wavelengths 
have revealed several important facts on the evolution of classical novae
during the outburst stage.
It shows that accretion is established and a disk is present early in the 
evolution
(as early as 645 days after the outburst). It also shows evidence that the irradiation of the secondary
stars by the hot stellar remnants (WDs) could be a common phenomenon that is
apparent in the outburst light curves of the first few years till after the stellar remnants cool off of the 
X-ray wavelengths. Only few novae have been detected during the SSS phase 
in the X-rays that lasts a few weeks up to several years after the outburst (eg. less than 9-10 years: Orio 2004) and such
observations are difficult to obtain due to the tight schedules of the X-ray satellites.
The evolution of the light variations as a result of the aspect changes of 
the irradiated secondary stars can yield additional indirect proof of the evolution
of the hot white dwarfs after the outburst and complement the X-ray observations if systematic long term ground-based
observations can be maintained in the optical wavelengths.

\acknowledgements
{The Authors acknowledge the American Association of Variable Stars (AAVSO) for 
providing the accumulated photometric data of V382 Vel from 1999-2003
(used to construct Figure 1). We thank S. Walker for sending us the 2000 data of V382 Vel.   
AR was partially supported by a research associate fellowship from Penn State University.
Finally, SB would like to acknowledge the support from the Turkish Academy of Sciences with
the TUBA-GEBIP (distinguished young scientist award) Fellowship. }

\begin{center}
\begin{table}
\footnotesize{ 
\label{1}
\caption{{\small The Time Table of the Observations (only Clear Filters were used)}}
\begin{tabular}{ccccc}
\hline
UT Date & Time of Start  & Run &  Number of &  \\
(ddmmyy) & (JD-2451000)  & Time (h) & Frames &
\\
\hline
300500 & 694.86480    & 3.6 & 368 &
\\
020600 & 697.84761   & 3.4 & 322 &
\\
070101 & 917.60610 & 0.4 & 117 &
\\
080101 & 918.60250 & 0.4 & 137 &
\\
110101 & 921.59740 & 0.7 & 239 &
\\
260201 & 966.83482  & 3.4 & 409 &
\\
270201 & 967.86559   & 6.1  & 553 &
\\
010301 & 969.86190   & 5.5  &  365 &
\\
030301 & 971.86294   & 8.6 &  405 &
\\
120301 & 980.89287  & 4.7 & 227 &
\\
150301 & 983.81037   & 4.8  &  216 &
\\
200301 & 988.86071   & 5.3  & 255 &
\\
220301 & 990.90068  & 5.4  & 265 &
\\
250301 & 993.80471  & 7.1  & 351 &
\\
080401 & 1007.80075  & 4.6  &  214 &
\\
140401  & 1013.82624   & 8.0 & 200 &
\\
150401 & 1014.00021 &  3.3 & 339 &
\\
210401 & 1020.86706 &  6.0 & 266 &
\\
260401 & 1025.84035 &  6.2 & 313 &
\\
160501 & 1045.78056 &  6.7 & 338 &
\\
310501 & 1060.77713 &  6.3  & 304 &
\\
030601 & 1063.76066 &  6.2  & 309 &
\\
010102 & 1275.94739 &  5.3 & 246 &
\\
020102 & 1276.89827  &  6.5  & 315 &
\\
060102 & 1280.96234 &  2.5  & 157 &
\\
070102 & 1281.86860 &  7.4  & 444 &
\\
200102 & 1294.87748 &   5.1  & 307 &
\\
210102 & 1295.86859 &   4.1 & 248 &
\\
220102 & 1296.86293  &  5.3  & 318 &
\\
\end{tabular}
}
\end{table}
\end{center}

\newpage

\figcaption{The AAVSO Lightcurve of V382 Vel obtained from the initial outburst
in May 1999 until 2003. The arrows indicate the observation times on Table 1.}

\figcaption{The figure presents the normalized differential light curve of V382 Vel
observed in 2000-2002. The observational dates are noted on each panel.
The data obtained in 2000 are taken with the 0.25 m telescope at the Wharemaru Observatory, New Zealand.
The rest of the data are taken with the 0.3 m Telescope, Auckland, New Zealand. Only clear filters are
used.  The epoch of the
observations are noted on the x axis. The error bars of the data points are about $0^m.03$,
$0^m.014$, and $0^m.013$ for the years 2000, 2001, and 2002, respectively. The scale of the x and y axes are 
fixed throughout the figure for better comparison.}

\figcaption{
The power spectrum of
V382 Vel, derived from the 2000 data set (top panel), the 2001 data set (middle panel), and the
2002 data set (bottom panel). The Scargle algorithm is used for the analyses.
The new period is indicated with P and its significant second harmonic
is also noted.
}

\figcaption{ The mean light curves of the 2001 and 2002 data sets, folded on the
period (P=0.146126(18) d) are presented at the top, and bottom panels, respectively. 
The middle panel shows the excess emission in phase once the binary period is prewhitened from the
light curve of 2001. The lines in the middle of the three panels show the average variation in the differential
magnitude of the three reference stars (used in the analysis) folded on the binary period (P=0.146126(18) d).  
The first data point in time (start mid-HJD in 2000)
is taken as the reference for the three mean light curves  
and a grouping (averaged over) of 25 phase bins (top), 20 phase bins (middle) and 20 phase bins
(bottom) are used for the folding process. }

\begin{center}

\begin{figure}
\plotone{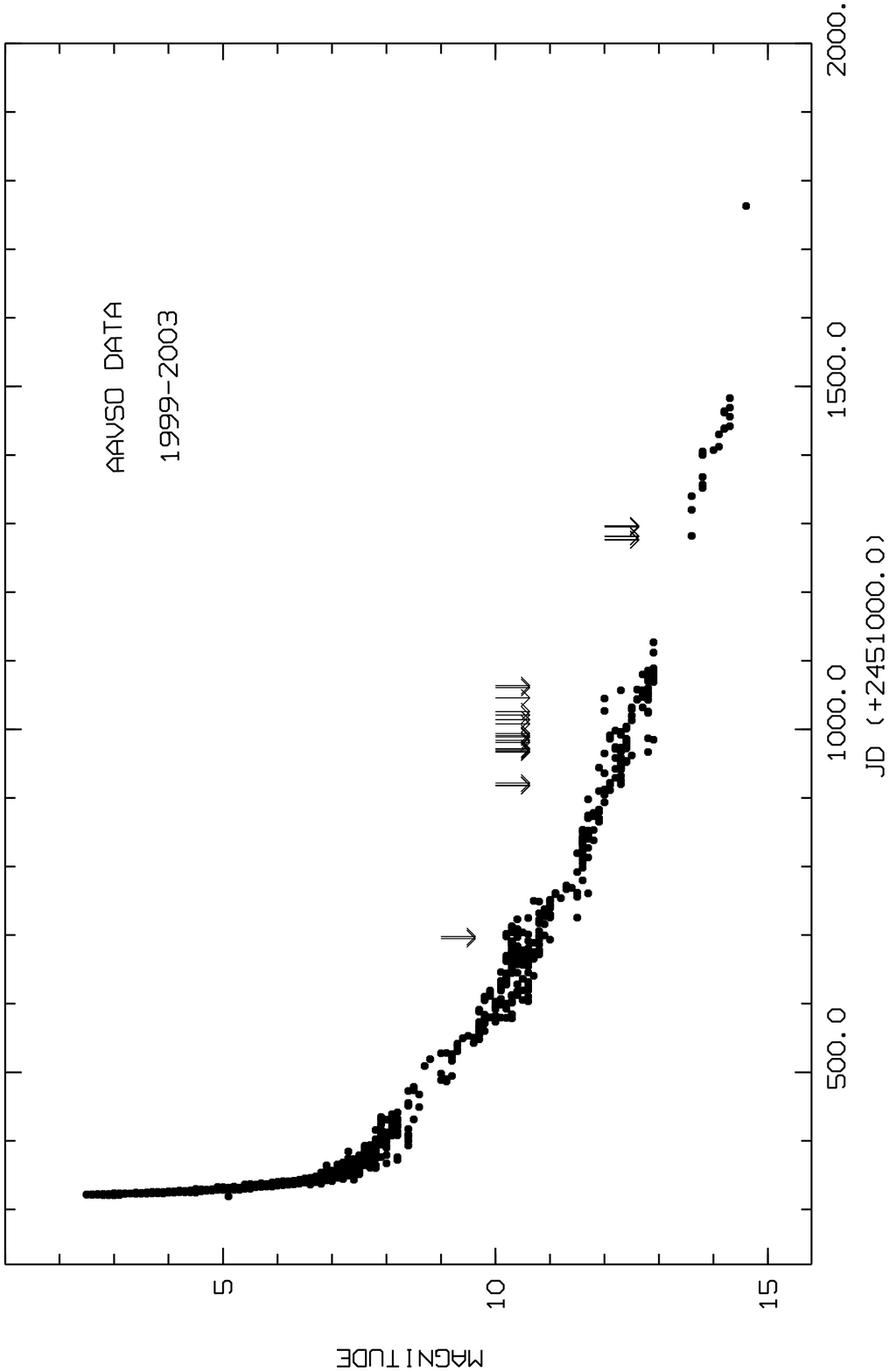}
\end{figure}                                                                                           

\begin{figure}
\plotone{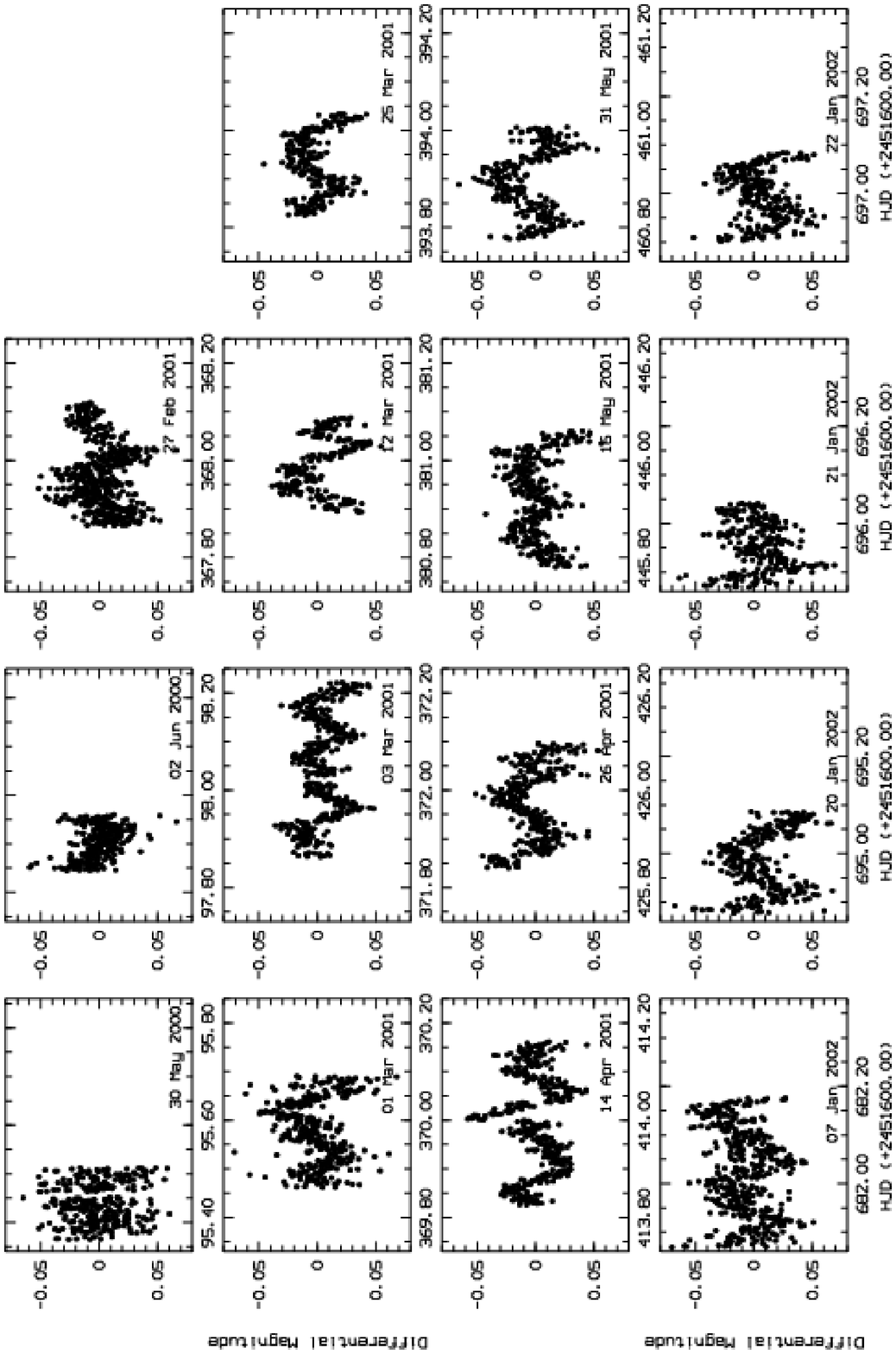}
\end{figure}        

\begin{figure}
\plotone{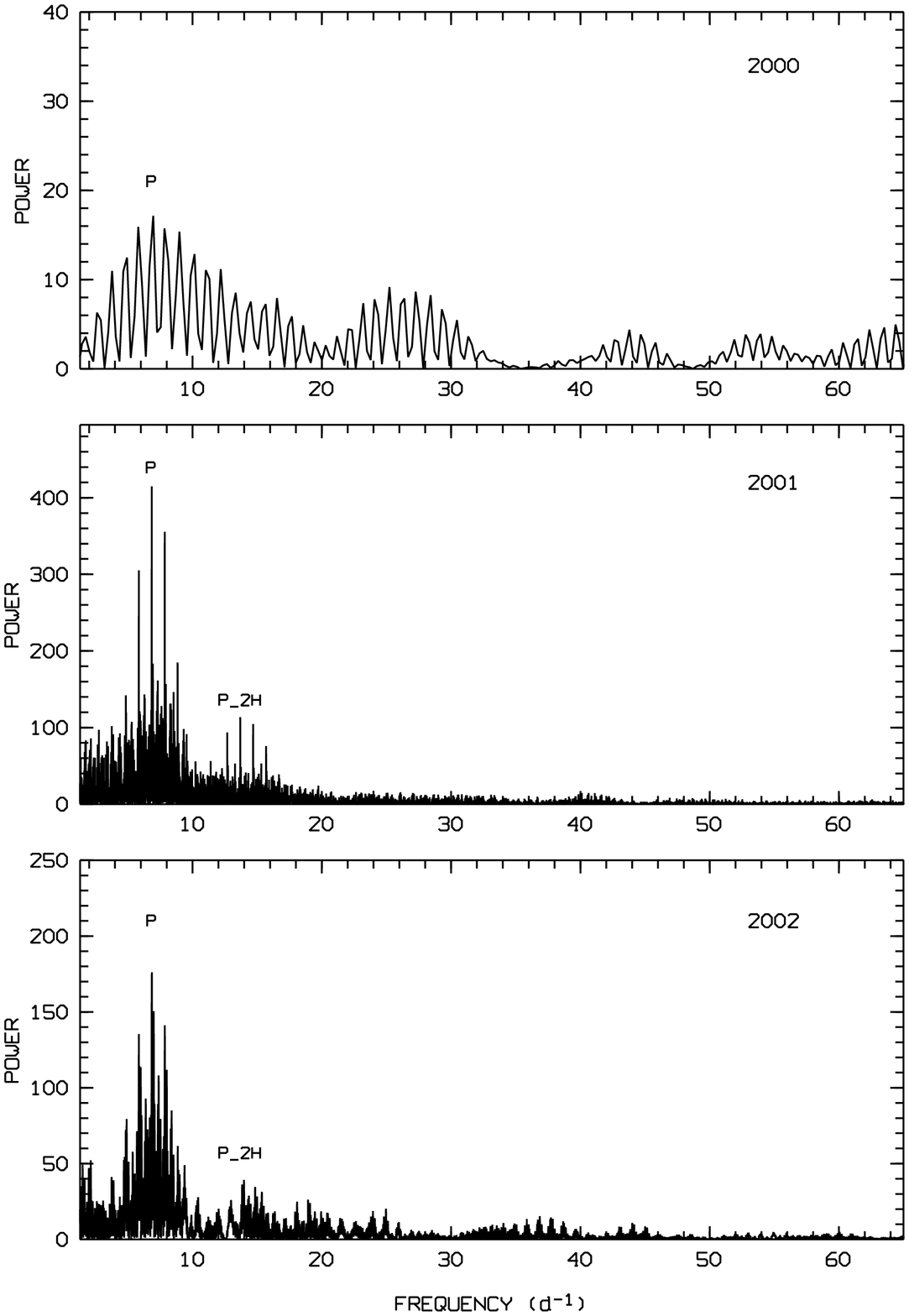}
\end{figure}

\begin{figure}
\plotone{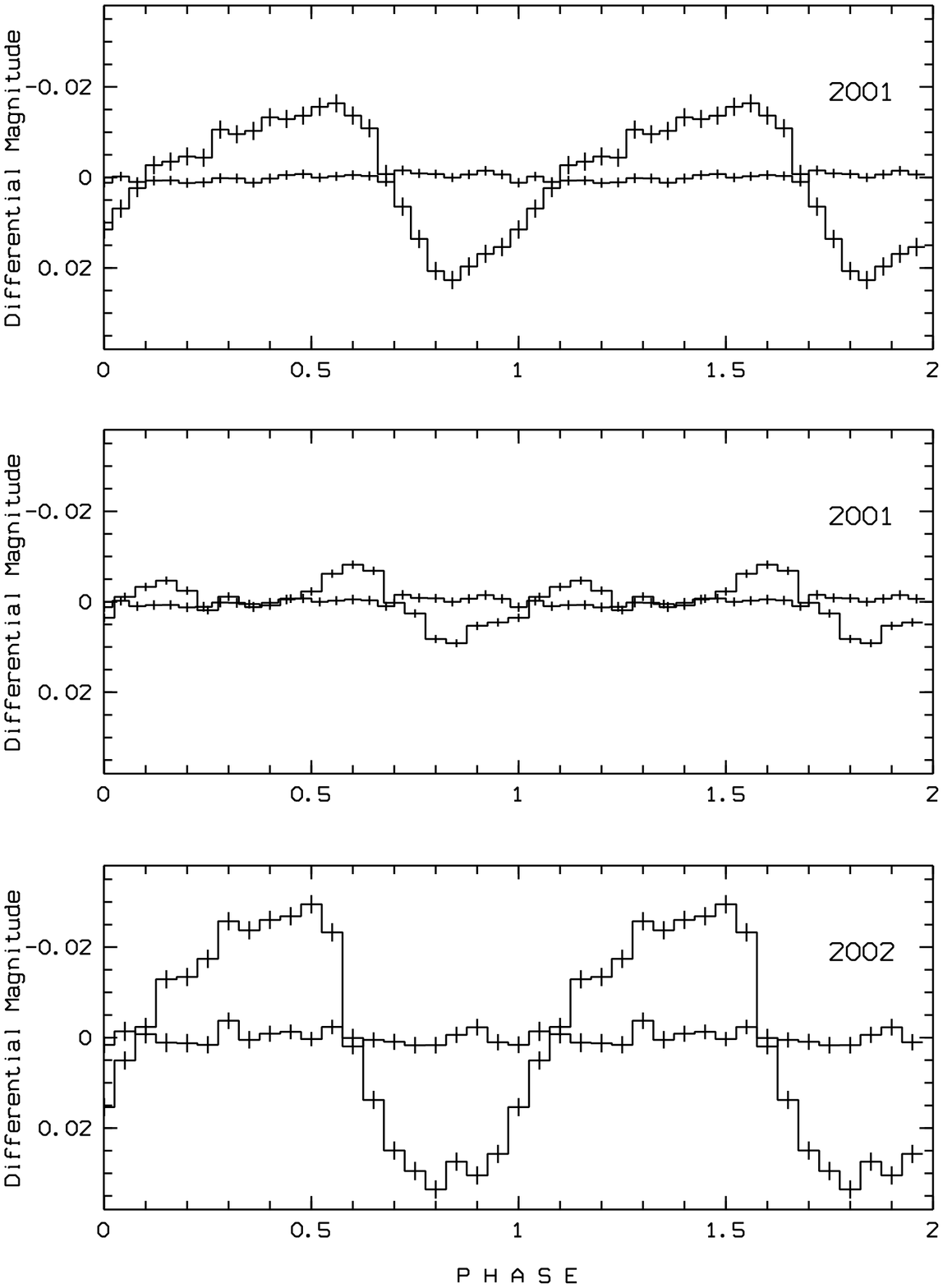}
\end{figure} 

\end{center}
\end{document}